\documentclass[prd,twocolumn,showpacs,nofootinbib,amsfonts,amssymb,amsmath]{revtex4-1}

\usepackage{graphicx}
\usepackage{simplewick}
\usepackage{epstopdf}
\usepackage{multirow}
\usepackage{cases}
\usepackage{color}

\def\be{\begin{equation}}
\def\ee{\end{equation}}
\def\bea{\begin{eqnarray}}
\def\eea{\end{eqnarray}}

\newcommand{\bear}{\begin{eqnarray}}

\newcommand{\eear}{\end{eqnarray}}

\newlength{\tskip}
\newbox\pippobox

\def\be{\begin{equation}}
\def\ee{\end{equation}}
\def\bea{\begin{eqnarray}}
\def\eea{\end{eqnarray}}

\def\a{\alpha}

\def\m{\mu}
\def\n{\nu}
\def\9{\nabla}

\def\b{\beta}

\def\dd{{\rm d}}

\def\p{\phi}

\def\nn{\nonumber}
\def\half{\frac12}
\def\le{\left}
\def\ri{\right}
\def\6{\partial}

\def\f{\frac}

\def\Mp{M_{pl}}
\def\tld{\tilde}
\def\0{(0)}
\def\half{\f{1}{2}}
\def\>{\rightarrow}

\begin{document}

\title{{\bf Parametrized modified gravity constraints after Planck}}

\author{Bin Hu$^{1}$}

\author{Michele Liguori$^{2,1}$}

\author{Nicola Bartolo$^{2,1}$}

\author{Sabino Matarrese$^{2,1}$}

\affiliation{$~^{1}$INFN, Sezione di Padova, via Marzolo 8, 35131 Padova, Italy
\\$~^{2}$Dipartimento di Fisica e Astronomia ``G. Galilei", Universit\`a degli Studi di Padova, via Marzolo 8, 35131 Padova, Italy}
\date{\today}

\begin{abstract}
We constrain $f(R)$ and chameleon-type modified gravity in the framework of the Berstchinger-Zukin parametrization using the recent released \textit{Planck} data, including both CMB temperature power spectrum and lensing potential power spectrum. Some other external data sets are included, such as baryon acoustic oscillations (BAO) measurements from the 6dFGS, SDSS DR7 and BOSS DR9 surveys, Hubble Space Telescope (HST) $H_0$ measurement and supernovae from Union2.1 compilation. We also use WMAP9yr data for consistency check and comparison. For $f(R)$ gravity, WMAP9yr results can only give quite a loose constraint on the modified gravity parameter $B_0$, which is related to the present value of the Compton wavelength of the extra scalar degree of freedom, 
$B_0<3.37$ at $95\% {\rm C.L.}$ We demonstrate that this constraint mainly comes from the late Integrated Sachs-Wolfe effect. With only {\it Planck} CMB temperature power-spectrum data, we can improve the WMAP9yr result by a factor $3.7$ ($B_0<0.91$ at $95\% {\rm C.L.}$). 
If the {\it Planck} lensing potential power-spectrum data are also taken into account, the constraint can be further strenghtened by a factor $5.1$ ($B_0<0.18$ at $95\% {\rm C.L.}$). This major improvement mainly comes from the small-scale lensing signal. 
Furthermore, BAO, HST and supernovae data could slightly improve the $B_0$ bound ($B_0<0.12$ at $95\% {\rm C.L.}$).
For the chameleon-type model, we find that the data set which we used cannot constrain the Compton wavelength $B_0$ and the potential index $s$ of chameleon field, but can give a tight constraint on the parameter $\beta_1=1.043^{+0.163}_{-0.104}$ at $95\% {\rm C.L.}$ ($\beta_1=1$ in general relativity), which accounts for the non-minimal coupling between the chameleon field and the matter component. In addition, we find that both modified gravity models we considered favor a relatively higher Hubble parameter than the concordance $\Lambda$CDM model in general relativity. 
\end{abstract}


\maketitle


\section{Introduction}
Cosmic acceleration can arise from either an exotic form of energy with negative pressure, 
referred to as ``dark energy", or a modification of gravity manifesting on large scales. 
As shown in \cite{Bertschinger:2006aw,Song:2006ej,Brax:2008hh}, at the the background level dark energy and modified gravity models are almost indistinguishable, 
hence one needs to investigate the perturbation dynamics. The studies of perturbation theory in modified gravity models, in principle, can be classified in two different frameworks: the parametrization approach and the non-parametrization method, 
such as the principal component analysis~\cite{Zhao:2009fn,Zhao:2010dz,Hojjati:2011xd}. 
In this paper we focus on the former.
There exist several phenomenological/theory-oriented parametrizations of modified gravity, such
as the Bertschinger-Zukin \cite{Bertschinger:2008zb} and the Brax-Davis-Li-Winther \cite{Brax:2012gr} parametrizations. 
These parametrizations are mainly suitable for the quasi-static regime, where the time evolution of the gravitational potentials is negligible compared with their spatial gradient. Furthermore, if we focus on the linear fluctuation dynamics, for which the equations in Fourier space can be reduced to simple algebraic relations, these techniques allow us to perform some analytic calculations which make the parametrization technically efficient. 
However, if we want to go beyond the quasi-static regime, while remaining in the linear perturbation framework, the parametrization of modified gravity becomes more complex. This is because on the largest scales, especially the super/near-horizon scales, the time evolution of the gravitational potentials is no longer negligible, the time derivative terms dominate the dynamical equations, which means that we need to solve some temporal ordinary differential equations.  
Actually, there exists some debate about the range of validity of the 
various parametrizations. For example, on one hand, as shown in \cite{Zuntz:2011aq}, using a parametrization with insufficient freedom significantly tightens the apparent theoretical constraints. 
On the other hand, for some specific modified gravity models some phenomenological parametrizations works quite well; for instance the authors of \cite{Hojjati:2012rf} recently demonstrated that for the 
small Compton wavelength case in the $f(R)$ model, the Bertschinger-Zukin parametrization \cite{Bertschinger:2008zb} is practically good enough for the current data analysis purpose.
This is because, on the scales larger than the Compton wavelength the deviation from general relativity is suppressed. Below the Compton scale the gravitational potential growth is enhanced and
the two metric potentials are no longer equal.
Consequently, for the small Compton wavelength case, whose value is less than current horizon size,
the most significant modifications w.r.t. general relativity occur in the sub-horizon regime.
In addition to the above explicit parametrizations, some quite generic frameworks to study different modified gravity scenarios have also been proposed, such as the Parameterized Post-Friedmann (PPF) formalism, including the Hu-Sawicki approach \cite{Hu:2007pj,Fang:2008sn}, its calibration version \cite{Lombriser:2013aj} and Baker-Ferreira-Skordis-Zuntz algorithm \cite{Baker:2011jy,Baker:2012zs}, and Effective Field Theory (EFT) approaches \cite{ArkaniHamed:2003uy,Gubitosi:2012hu,Bloomfield:2012ff,Creminelli:2006xe,Cheung:2007st,Creminelli:2008wc,Gleyzes:2013ooa,Bloomfield:2013efa}. 

On the observational point of view, many windows have been proposed to constrain modified gravity models, such as the Integrated Sachs-Wolfe (ISW) effect~\cite{Sachs:1967er} in Cosmic Microwave Background (CMB) anisotropies, including CMB power spectrum \cite{Zhang:2005vt,Ho:2008bz,Daniel:2010ky,Bean:2010zq,Zhao:2010dz,Marchini:2013lpp}, CMB ISW-Lensing bispectrum \cite{DiValentino:2012yg,Hu:2012td}, baryon acoustic oscillations (BAO) measurements \cite{Yamamoto:2006yv,Bean:2010zq,Said:2013jxa}, the galaxy-ISW cross correlation~\cite{Song:2007da,Giannantonio:2009gi,Bean:2010zq,Dossett:2011zp}, cluster abundance~\cite{Jain:2007yk,Lombriser:2010mp,Ferraro:2010gh,Lombriser:2013wta}, peculiar velocity~\cite{Li:2012by,Asaba:2013xql}, redshift-space distortions~\cite{Jennings:2012pt,Raccanelli:2012gt}, weak-lensing~\cite{Zhang:2007nk,Hirata:2008cb,Zhao:2008bn,Reyes:2010tr,Zhao:2010dz,Daniel:2010ky,
Bean:2010zq,Dossett:2011zp,Laszlo:2011sv,Tereno:2010dt,Asaba:2013xql,Thomas:2008tp,Cai:2011wj}, $21$cm lines~\cite{Wang:2010ug,Hall:2012wd}, matter power spectrum and bispectrum~\cite{Yamamoto:2010ie,He:2012wq,GilMarin:2011xq,Bartolo:2013ws}. 
In addition, recently some N-body simulation algorithms in modified gravity models have been developed~\cite{Zhao:2010qy,Li:2010zw,He:2013vwa}. As shown in \cite{Song:2007da,Lombriser:2010mp,He:2012wq}, with WMAP resolution the modification effects on the CMB mainly come from the ISW effect, which becomes prominent on the largest scales. However, due to the unavoidable cosmic variance on large scales, the constraints from these effects are not significant. On the other hand, since the typical modification scales are in the sub-horizon regime, several studies show that the most stringent constraints come from the large-scale structure data sets. For example, the strongest current constraint on $f(R)$ gravity ($B_0<1.1\times 10^{-3}\;,95\%{\rm C.L.}$)~\cite{Lombriser:2010mp} is obtained through cluster abundance data sets. Various previous results show that the main constraint on modified gravity comes from galaxy or cluster scales which corresponds to the multipole range $l\gtrsim 500$ in CMB data, where 
lensing effect is no longer negligible. The recent release of {\it Planck} data \cite{Ade:2013ktc} provides us with a fruitful late-time information both on the ISW and lensing scales, which is encoded in the CMB temperature power-spectrum 
\cite{Planck:2013kta} and lensing potential power-spectrum \cite{Ade:2013tyw} and CMB temperature ISW-Lensing bispectrum \cite{Ade:2013ydc,Ade:2013dsi}. 
The full sky lensing potential map has been firstly measured and the significance of the amplitude of the lensing potential power-spectrum arrives at the $25\sigma$ level. The ISW-Lensing bispectrum is also firstly detected with nearly $3\sigma$ significance. Furthermore, through the lensing potential reconstruction and the ISW-Lensing bispectrum, the ISW effect is also firstly detected via the CMB itself. 
All in all, with its high resolution the {\it Planck} mission provides us with fruitful information about the universe late-time acceleration. 
For example, the authors of \cite{Li:2013nwa} shows that the joint analysis of {\it Planck} and BAO data could greatly improve the Brans-Dicke parameter $\omega$ constraint. 
Further new constraint results related with modified gravity/dark energy can be found in \cite{Li:2013dha,Salvatelli:2013wra,Pettorino:2013oxa,Marchini:2013oya,He:2013qha}. 

Due to these considerations, in this paper we investigate the power of the {\it Planck} data sets in constraining modified gravity scenarios. In order to break the parameter degeneracies, apart from {\it Planck} data sets, we also use some external astrophysical data sets, such as BAO measurements from the 6dFGS, SDSS DR7 and BOSS DR9 surveys, $H_0$ from HST measurement and supernovae from Union2.1 compilation. We also use WMAP9yr data for consistency check and comparison.
Because of the simplicity of the Bertschinger-Zukin parametrization, in this paper we study the modified gravity theory through this method.

\section{Bertschinger-Zukin parametrization}

As pointed out in \cite{Brax:2012gr}, a large class of modified gravity theories, e.g., chameleon \cite{Mota:2006fz,Brax:2004qh}, symmetron \cite{Pietroni:2005pv,Olive:2007aj,Hinterbichler:2010es} and dilaton \cite{Brax:2010gi} models can be characterized by the mass of a suitable scalar field and the coupling between the scalar field and baryonic/dark matter components. In the Einstein frame, where the gravitational sector is the standard Einstein-Hilbert action, the scalar field is exponentially coupled with the matter sector
\bea
S_{\rm E}&=&\int\dd^4x\sqrt{-\tld g}\le[\f{\Mp^2}{2}\tld R-\half\tld g^{\m\n}(\tld\9_{\m}\p)(\tld\9_{\n}\p)-V(\p)\ri]\nn\\
&+& S_i(\chi_i,e^{-\kappa \a_i(\p)\tld g_{\m\n}})\;,\eea
where the Einstein frame metric $\tld g_{\m\n}$ is related to the Jordan frame one $g_{\m\n}$ through a conformal transformation
\be
\tld g_{\m\n}=e^{\kappa \a_i(\p)}g_{\m\n}\;,\ee
and $\chi_i$ denotes the matter components. 

Inspired by some nice properties in the quasi-static regime of $f(R)$ model, Bertschinger and Zukin in \cite{Bertschinger:2008zb} first write the two gravitational potentials in the conformal Newtonian gauge \footnote{We take the convention that $\dd s^2 = -(1+2\Psi)\dd t^2 + a^2(1-2\Phi)\dd x^2$.} in terms of two observation-related variables, the time- and scale-dependent Newton's constant $G \mu(a,k)$ and the so-called gravitational slip $\gamma(a,k)$
\bea
\label{BZ}
k^2\Psi &=& -4\pi Ga^2\mu(a,k)\rho\Delta\;,\\
\label{BZ1.2}
\frac{\Phi}{\Psi}&=&\gamma(a,k)\;,
\eea
where $G$ is the Newton's constant in the laboratory.
The corresponding Einstein-Boltzmann solver named \emph{MGCAMB} is implemented 
in \cite{Zhao:2008bn,Hojjati:2011ix}. In this paper, we implement the same algorithm in the new version of \emph{CAMB} \cite{Lewis:1999bs} which is compatible with the {\it Planck} likelihood. 

In the following sections, we will study $f(R)$ gravity and the quite general chameleon-type model in the framework of the Bertschinger-Zukin parameterized modified gravity method, by using the {\it Planck} \cite{Planck:2013kta,Ade:2013tyw} WMAP9yr \cite{Bennett:2012fp,Hinshaw:2012aka} and some external astrophysical data.

\subsection{$f(R)$ model}
\begin{figure}[ht]
\begin{center}
  \includegraphics[width=0.5\textwidth]{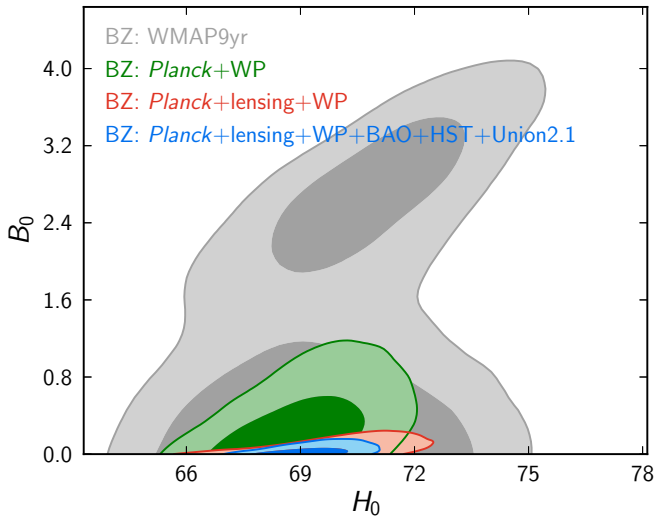}
  \caption{\label{B0_H0} Two-dimensional contour diagram of $B_0$ and $H_0$.
  The appearance of the upper dark gray area is due to the non-linear dependence 
  of the ISW effect on $B_0$.}
\end{center}
\end{figure}

\begin{figure}[ht]
\begin{center}
  \includegraphics[width=0.5\textwidth]{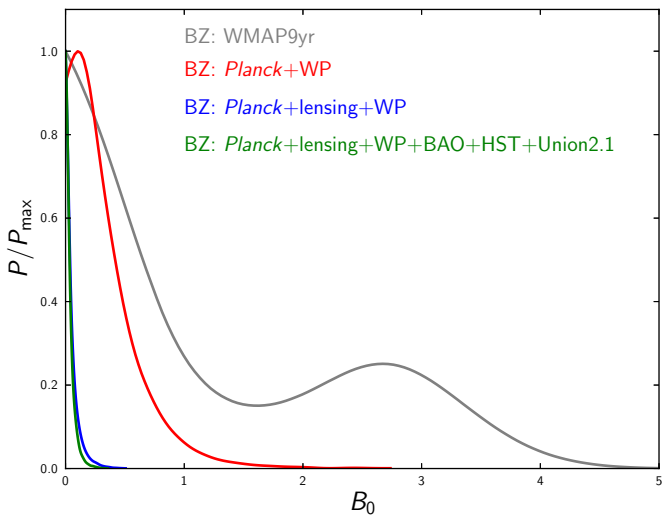}
  \caption{\label{B0} The likelihood of $B_0$. The second peak in the gray curve is due to 
  the non-linear dependence of ISW effect on $B_0$.}
\end{center}
\end{figure}

\begin{figure}[ht]
\begin{center}
  \includegraphics[width=0.5\textwidth]{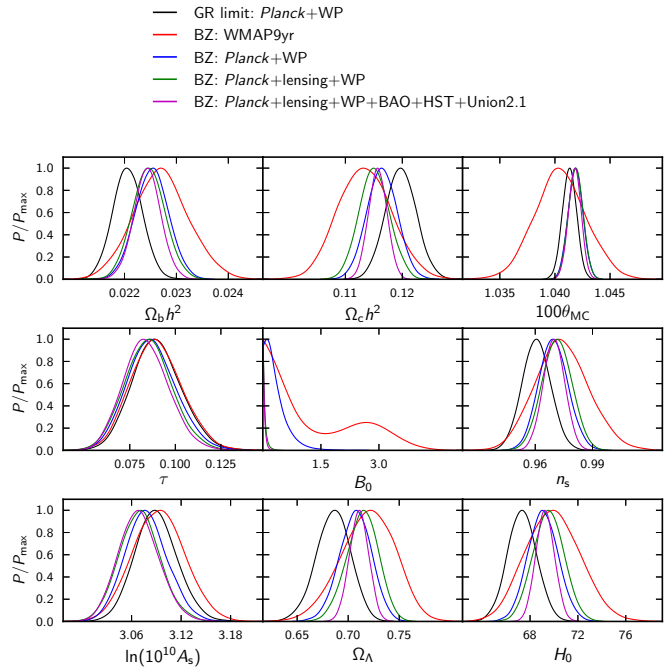}
  \caption{\label{MG_1d} Full set of parameter likelihoods in $f(R)$ gravity.}
\end{center}
\end{figure}
Due to the simplicity of its Lagrangian, $f(R)$ gravity obtained a lot of attention,
(see the recent review \cite{DeFelice:2010aj} and references therein) especially as an illustration of the chameleon mechanism. Besides the simplicity of the structure of this theory, there exist two more reasons for the interest it attracted. 
One is that the form of the function $f(R)$ can be engineered to exactly mimic {\it any} background history via a one-parameter family of solutions~\cite{Song:2006ej}. The second reason is that $f(R)$ gravity can slightly better fit than ﬂat $\Lambda$CDM, which can be attributed to the lowering of the temperature anisotropy power spectrum at small $l$ regime~\cite{Lombriser:2010mp}. In this paper we consider the class of $f(R)$ gravity models which can mimic a $\Lambda$CDM background.

Because of the higher order derivative nature of $f(R)$ gravity, there exist a scalar degree of freedom, named scalaron
$f_R\equiv\dd f/\dd R$ with mass
\be\label{fR_mass}
m^2_{f_R}\equiv\frac{\6^2V_{\rm eff}}{\6 f^2_{R}}=\frac{1}{3}\left(\frac{1+f_R}{f_{RR}}-R\right)\;.\ee
Then the Compton wavelength of the scalaron reads
\be\label{comp_wav1}
\lambda_{f_R}\equiv m^{-1}_{f_R}\;.\ee
Usually, it is convenient to use the dimensionless Compton wavelength
\be\label{comp_wav2}
B\equiv\f{f_{RR}}{1+f_R}R'\f{H}{H'}\;,\ee
with $f_{RR}=\dd^2 f/\dd R^2$ and ${~}'=\dd /\dd \ln a$. 

In the Bertschinger-Zukin parametrization \cite{Bertschinger:2008zb}, the explicit expressions of the functions $\mu(a,k)$ and $\gamma(a,k)$ for $f(R)$ gravity read
\bea
\label{BZ2.1}
\mu(a,k) &=& \frac{1+\frac{4}{3}\lambda_1^2k^2a^4}{1+\lambda_1^2k^2a^4}\;,\\
\label{BZ2.2}
\gamma(a,k) &=& \frac{1+\frac{2}{3}\lambda_1^2k^2a^4}{1+\frac{4}{3}\lambda_1^2k^2a^4}\;, 
\eea
based on the quasi-static approximation. 
The above parametrization is improved by Giannantonio {\it et. al.} in~\cite{Giannantonio:2009gi} to take the ISW effect into account through some empirical formula
\be
\label{BZ3}
\mu(a,k) = \frac{1}{1-1.4\cdot 10^{-8}|\lambda_1|^2a^3}\frac{1+\frac{4}{3}\lambda_1^2k^2a^4}{1+\lambda_1^2k^2a^4}\;.
\ee
Due to this reason, in our numerical calculation we use (\ref{BZ3}) instead of the original expression (\ref{BZ2.1}).

Through a few simple computations, one can easily find that $\lambda_1$ is nothing but the present Compton wavelength $\lambda_1^2=B_0c^2/(2H_0^2)$. 
Remember that Song {\it et. al.} in \cite{Song:2006ej} pointed out that there exists a one-parameter family solution in $f(R)$ gravity which could mimic any background evolution. 
Conventionally, we choose this one-parameter family labeled by the Compton wavelength at present $B_0$ or $\lambda_1^2$ in the Bertschinger-Zukin parametrization. 
Given the above analysis, we can see that in $f(R)$ gravity, compared with the concordance $\Lambda$CDM model, there is only one extra parameter, $B_0$, 
which makes the effects of gravitational modification quite manifest. 

\subsection{Chameleon-type model}
\begin{figure}[ht]
\begin{center}
  \includegraphics[width=0.5\textwidth]{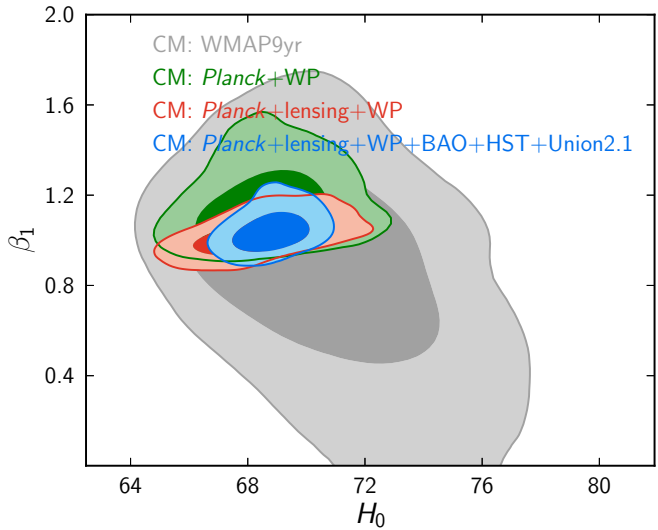}
  \caption{\label{beta1_H0} Two-dimensional contour of $\beta_1$ and $H_0$.}
\end{center}
\end{figure}

\begin{figure}[ht]
\begin{center}
  \includegraphics[width=0.5\textwidth]{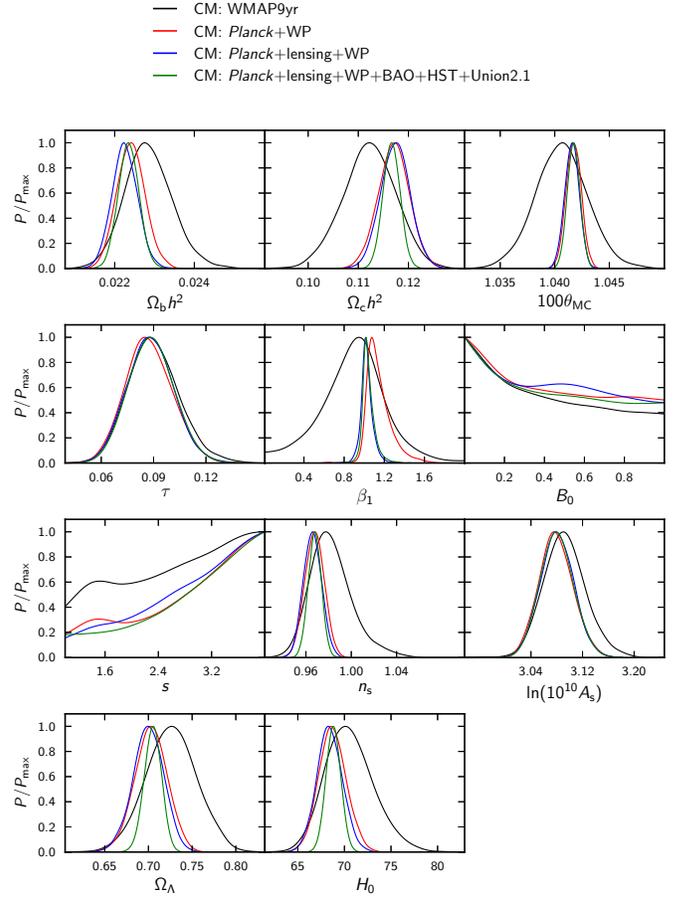}
  \caption{\label{CM_1d} Full set of parameter likelihoods in chameleon-type model.}
\end{center}
\end{figure}
The chameleon models \cite{Mota:2006fz,Brax:2004qh} are
characterised by a runaway potential and a nearly constant coupling $\a$.
Since the $f(R)$ model can be seen as a specific chameleon model, it is straightforward to
generalize the Bertschinger-Zukin parametrization for $f(R)$ gravity (\ref{BZ2.1}) and (\ref{BZ2.2}) into
\bea
\label{BZ4.1}
\mu(a,k) &=& \frac{1+\beta_1\lambda_1^2k^2a^s}{1+\lambda_1^2k^2a^s}\;,\\
\label{BZ4.2}
\gamma(a,k) &=& \frac{1+\beta_2\lambda_2^2k^2a^4}{1+\lambda_2^2k^2a^4}\;,
\eea
where the parameters need to satisfy the following relation
\be
\label{relation}
\beta_1=\frac{\lambda_2^2}{\lambda_1^2}=2-\beta_2\frac{\lambda_2^2}{\lambda_1^2}\;,
\ee
and $1\leq s \leq 4$. Via the above constraints the number of free parameters can be
reduced to $3$, usually, one choose them as ($s,\beta_1,\lambda_1$). In \cite{Giannantonio:2009gi,Hojjati:2011ix} this kind of parametrization is called Yukawa-type models, due to the Yukawa-type interaction between dark matter particles.

Because of the non-minimal coupling, the dynamics of the scalar field is determined jointly by the scalar field and the matter component, for example, the effective potential of the scalar field is defined by
\be
V_{\rm eff}(\p) = V(\p) + \bar\rho_i e^{\kappa \a_i(\p)}\;,\ee
which gives an effective mass of chameleon field
\be
m^2 = V''_{\rm eff} = V'' - \kappa(\a'' + \a^{'2})V'\;,\ee
where primes denote differentiation w.r.t. the field.
For simplicity, here we assume that the chameleon field couples to all the matter components uniformly.
Following some calculations as in \cite{Brax:2004qh,Zhao:2008bn}, we can 
obtain the following relations
\bea
\a^{1+s/2} &=& \f{m_0}{m}\;,\nn\\
\lambda_1^2 &=& \f{1}{m_0^2}\;,\nn\\
\lambda_2^2 &=& \f{1}{m_0^2}\le(1+\f{\a^{'2}}{2}\ri)\;,\nn\\
\b_1 &=& 1+ \f{\a^{'2}}{2}\;,\nn\\
\b_2 &=& \f{2-\a^{'2}}{2+\a^{'2}}\;,\eea
where $m_0$ is the chameleon effective mass at present.
Furthermore, for the inverse power-law potential $V(\p)\propto \p^{-n}$ case, with $n>0$, we have
\be
n=\f{4-s}{s-1}\;.\ee
Here $\lambda_1$ can be replaced with the conventional parameter $B_0$, with the same expression in $f(R)$ model, namely $\lambda_1^2=B_0c^2/(2H_0^2)$.
Through the above relations, we can easily see that the parameters $\beta_1$, $B_0$ and $s$ correspond to the non-minimally coupling between chameleon field and matter sector, 
the relative Compton wavelength of chameleon field and the potential index of chameleon field, respectively.
 Moreover, the general relativity limit corresponds to $\beta_1=1,B_0=0,s=4$ \cite{Zhao:2008bn}.

\section{Data analysis methodology}
\begin{figure}[ht]
\begin{center}
  \includegraphics[width=0.5\textwidth]{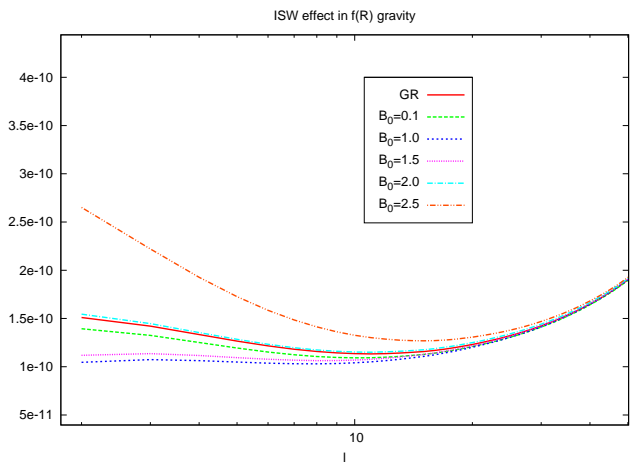}
  \caption{\label{isw} The non-linear dependence of the ISW effect on $B_0$ in $f(R)$ gravity.}
\end{center}
\end{figure}
The purpose of this paper is to test possible deviations from general relativity on various cosmic scales by using the recent {\it Planck} data, including both the 
CMB temperature and lensing potential power-specta and also some external astrophysical data sets.  
In the following section, we will briefly review the {\it Planck} likelihood and data set which we used in this work. 

The total {\it Planck} CMB temperature power-spectrum likelihood is divided into low-$l$ ($l<50$) and high-$l$ ($l\geq 50$) parts. 
This is because the central limit theorem ensures that the distribution of CMB angular power spectrum $C_l$ in the high-$l$ regime can be well approximated by a Gaussian statistics. However, for the low-$l$ part the $C_l$ distribution is non-Gaussian. 
For these reasons the {\it Planck} team adopts two different methodologies to build the likelihood. In detail,
for the low-$l$ part, the likelihood exploits all {\it Planck} frequency channels
from $30$ to $353$ GHz, separating the cosmological CMB signal from diffuse Galactic foregrounds through a physically motivated Bayesian component separation technique. 
For the high-$l$ part, the {\it Planck} team employ a correlated Gaussian likelihood approximation, based on a fine-grained set of angular cross-spectra derived from multiple detector combinations between the $100$, $143$, and $217$ GHz frequency channels, marginalizing over power-spectrum foreground templates. In order to break the 
well-known parameter degeneracy between the reionization optical depth $\tau$ and the scalar index $n_s$, the {\it Planck} team assumed the low-$l$ WMAP polarization likelihood (WP).
Apart from the CMB power-spectrum, the first {\it Planck} data release provides for the first time a full-sky lensing potential map,  
by using the $100$, $143$, and $217$ GHz frequency bands with an overall significance greater than
$25\sigma$. As we know, the lensing potential distribution follows that of the large-scale structures which form and grow mainly in the late-time universe. Thus, this map carries fruitful information about dark energy/modified gravity 
in this period. Hence, we expect that the lensing potential power-spectrum could provide us with a stringent constraint on deviations from general relativity. 

Given the above considerations, we perform our parameter estimation algorithms by using two different data sets from the {\it Planck} mission, namely the {\it Planck} CMB power-spectrum \cite{Planck:2013kta} and lensing potential power-spectrum \cite{Ade:2013tyw}. 
In order to compare with the previous WMAP results, we also do the same analysis by using the WMAP9yr data \cite{Hinshaw:2012aka}. 
Furthermore, in order to break the parameter degeneracies we also use some other external data sets, including baryon acoustic oscillations (BAO) measurements from the 6dFGS \cite{Beutler:2011hx}, SDSS DR7 \cite{Percival:2009xn}, and BOSS DR9 \cite{Anderson:2012sa} surveys, Hubble Space Telescope (HST) Key Project \cite{Freedman:2000cf} $H_0$ measurement and supernovae from Union2.1 compilation \cite{Suzuki:2011hu}. For BAO data sets, we use three redshift survey:
the 6dF Galaxy Survey measurement at $z= 0.1$, the reanalyzed SDSS-DR7 BAO measurement \cite{Padmanabhan:2012hf} at effective redshift $z_{\rm eff}=0.35$, and the BOSS-DR9 measurement at $z_{\rm eff}=0.2$ and $z_{\rm eff}=0.35$. For the direct measurement of the Hubble constant, we use the result $H_0 = 73.8\pm 2.4 {\rm km s^{-1} Mpc^{-1}}$ \cite{Riess:2011yx}, which comes from the supernova magnitude-redshift relation calibrated by the HST observations of Cepheid variables in the host galaxies of eight SNe Ia. For supernovae, we use the Union2.1 compilation, consisting of 580 SNe, calibrated by the SALT2 light-curve fitting model.

\begin{table}
\begin{tabular}{c|c|c}
\hline\hline
Parameter & \multicolumn {2}{c}{Range (min, max)}  \\
\hline
$\Omega_b h^2$ & \multicolumn {2}{c}{$(0.005,
0.100)$} \\
$\Omega_c h^2$ &  \multicolumn {2}{c}{$(0.01, 0.99)$}  \\
$100\vartheta_*$ &
\multicolumn {2}{c}{$(0.5, 10.0)$ }  \\
$\tau$ & \multicolumn {2}{c}{$(0.01, 0.80)$ } \\
$n_s$ &  \multicolumn {2}{c}{$(0.5, 1.5)$} \\
$\ln (10^{10} A_s^2)$ &  \multicolumn {2}{c}{$(2.7, 4.0)$} \\
\hline
MG parameters & $f(R)$ & Chameleon \\
 \hline
$ B_0   $  &  $(0, 10)$ & $(0, 1)$ \\
$\beta_1$   & $ 4/3 $ & $(0.001, 2)$ \\
$s$       & $ 4 $  & $(1, 4)$ \\
\hline\hline
\end{tabular}
\caption{List of the parameters used in the Monte Carlo sampling.}
 \label{tab:parameters}
\end{table}

As previously stated, we implement the same algorithms of \emph{MGCAMB} \cite{Zhao:2008bn,Hojjati:2011ix} in the new version of \emph{CAMB} \cite{Lewis:1999bs}, which is compatible with the {\it Planck} likelihood. 
We sample the cosmological parameter space, which can be read in Tab.\ref{tab:parameters}, with a Markov Chain Monte Carlo (MCMC) method with the publicly available code \texttt{CosmoMC} \cite{Lewis:2002ah}.

\begin{figure}[ht]
\begin{center}
  \includegraphics[width=0.5\textwidth]{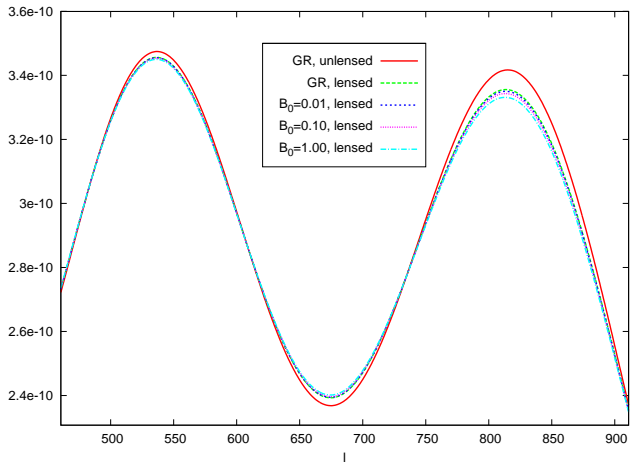}
  \caption{\label{3rdpeak} The second and third peaks in $f(R)$ gravity. The larger $B_0$ is the lower the third peak is.}
\end{center}
\end{figure}

\section{Results and Discussion}
As a first step we checked the reliability of the code in the general relativity limit ($B_0=0$ for $f(R)$ gravity case, $B_0=0,\beta_1=1,s=4$ for a chameleon-type model). We find that our results are in quite good agreement with the {\it Planck} results \cite{Ade:2013zuv}. 
Here we show our consistency check for the $f(R)$ case explicitly in Tab. \ref{Tab2}.

\setlength\tabcolsep{1pt}
\begin{table*}[htb!]
\footnotesize
\begin{center}
\begin{tabular}{|l||c c||c c|c c|c c|c c|}
\hline
\hline
&  \multicolumn{2}{|c||}{{\bf GR limit:\textit{Planck}+WP}} & \multicolumn{2}{|c|}{{\bf BZ:WMAP9yr}} & \multicolumn{2}{|c|}{{\bf \textit{Planck}+WP}} & \multicolumn{2}{|c|}{{\bf +lensing}} & \multicolumn{2}{|c|}{{\bf +BAO+HST+Union2.1}}\\ \hline
Parameters & Best fit & $68\%$ limit & Best fit & $68\%$ limit &  Best fit & $68\%$ limit & Best fit & $68\%$ limit &  Best fit & $68\%$ limit\\ \hline
$\Omega_b h^2$ &.02266 & .02206$\pm$.00028 & .02270 & .02271$\pm$.00052 &  .02250 & .02253$\pm$.00031 & .02227  & .02247$\pm$.00031 & .02232 & .02244$\pm$.00026\\
$\Omega_c h^2$ &.1201 & .1198$\pm$.0027 & .1147 & .1134$\pm$.0046 & .1178 & .1164$\pm$.0026 & .1173 & .1151$\pm$.0026 & .1180 & .1157$\pm$.0016\\
$100\theta$ & 1.04151 & 1.04132$\pm$.00063 & 1.0410 & 1.0405$\pm$.0023 & 1.0420 & 1.04190$\pm$.00065 & 1.0418 & 1.0419$\pm$.00064 & 1.0413 & 1.0418$\pm$.00057\\
$\tau$ &.083 & .090$\pm$.013 & .086 & .090$\pm$.014 & .077 & .087$\pm$.013 & .103 & .085$\pm$.013 & .092 & .084$\pm$.012\\
$n_s$ &.9601 & .9607$\pm$.0073 & .973 & .974$\pm$.014 & .967 & .970$\pm$.0075 & .970 & .971$\pm$.0076 & .965 & .970$\pm$.0056\\
$\log(10^{10} A_s)$ & 3.077 & 3.090$\pm$.025 & 3.092 & 3.093$\pm$.031 & 3.063 & 3.078$\pm$.025 & 3.111 & 3.070$\pm$.024 & 3.091 & 3.069$\pm$.024\\
$B_0$ & ------ & ------ & .015 & $<$1.94(3.37) & .121 & $<$.38(.91) & .023 & $<$.054(.18) & .0044 & $<$.041(.12)\\
\hline
$\Omega_{\rm \Lambda}$ & .684  & .685$\pm$.016 & .715 & .719$\pm$.026 & .701 & .707$\pm$.015 & .702 & .713$\pm$.015 & .697 & .711$\pm$.0092\\
$H_0 [\mathrm{km}/\mathrm{s}/\mathrm{Mpc}]$ & 67.25 & 67.34$\pm$1.19 & 69.59 & 69.92$\pm$2.23 & 68.64 & 69.09$\pm$1.24 & 68.56 & 69.54$\pm$1.26 & 68.15 & 69.27$\pm$.76 \\
\hline
\hline
$\chi^2_{\rm min}/2$ &\multicolumn{2}{|c||}{4902.724} & \multicolumn{2}{|c|}{3779.201} & \multicolumn{2}{|c|}{4900.427} & \multicolumn{2}{|c|}{4907.413} & \multicolumn{2}{|c|}{4975.704}\\
\hline
\hline
\end{tabular}
\caption{Best-fit values and $68\%$ confidence limits for $f(R)$ gravity (and $95\%$ confidence limits in parenthesis for $B_0$).
The first column shows the consistency check of the code in the general relativity limit.}
\label{Tab2}
\end{center}
\end{table*}

\setlength\tabcolsep{1pt}
\begin{table*}[htb!]
\begin{center}
\begin{tabular}{|l||c c|c c|c c|c c|}
\hline
\hline
&  \multicolumn{2}{|c|}{{\bf CM:~WMAP9yr}} & \multicolumn{2}{|c|}{{\bf CM:~\textit{Planck}+WP}} & \multicolumn{2}{|c|}{{\bf +lensing}}  & \multicolumn{2}{|c|}{{\bf +BAO+HST+Union2.1}}\\ \hline
Parameters & Best fit & $68\%$ limit &  Best fit & $68\%$ limit & Best fit & $68\%$ limit & Best fit & $68\%$ limit \\ \hline
$\Omega_b h^2$ & .02279 & .02286$\pm$.00059 &  .02256 & .02241$\pm$.00035 & .02226  & .02225$\pm$.00032 & .02240 & .02235$\pm$.00026\\
$\Omega_c h^2$ & .1184 & .1122$\pm$.0052 & .1168 & .1171$\pm$.0031 & .1162 & .1174$\pm$.0029 & .1168 & .1166$\pm$.0017\\
$100\theta$ & 1.0391 & 1.0406$\pm$.0024 & 1.04158 & 1.04174$\pm$.00068 & 1.04183 & 1.04158$\pm$.00065 & 1.04159 & 1.04173$\pm$.00057\\
$\tau$ & .092 & .090$\pm$.015 & .088 & .087$\pm$.013 & .089 & .088$\pm$.013 & .090 & .089$\pm$.013\\
$n_s$ & .9879 & .9825$\pm$.019 & .9686 & .9676$\pm$.0084 & .9659 & .9658$\pm$.0079 & .9698 & .9678$\pm$.0057\\
$\log(10^{10} A_s)$ & 3.131 & 3.092$\pm$.033 & 3.082 & 3.079$\pm$.026 & 3.079 & 3.081$\pm$.025 & 3.085 & 3.080$\pm$.025\\
$\beta_1$ & 0.954 & $0.893^{+0.647}_{-0.695}$ & 1.127 & $1.148^{+0.274}_{-0.194}$ & 1.033 & $1.027^{+0.140}_{-0.114}$ & 1.020 & $1.043^{+0.163}_{-0.104}$\\
$B_0$ & 0.496 & ------ & 0.849 & ------ & 0.473 & ------ & 0.079 & ------\\
$s$ & 1.143 & ------ & 3.398 & ------ & 3.152 & ------ & 3.635 & ------\\
\hline
$\Omega_{\rm \Lambda}$ & .691 & .726$\pm$.029 & .705 & .703$\pm$.018 & .701 & .700$\pm$.017 & .704 & .705$\pm$.0098\\
$H_0 [\mathrm{km}/\mathrm{s}/\mathrm{Mpc}]$ & 67.64 & 70.58$\pm$2.59 & 68.88 & 68.73$\pm$1.46 & 68.93 & 68.43$\pm$1.36 & 68.75 & 68.82$\pm$.78\\
\hline
\hline
$\chi^2_{\rm min}/2$ & \multicolumn{2}{|c|}{3778.939} & \multicolumn{2}{|c|}{4900.274} & \multicolumn{2}{|c|}{4907.445} & \multicolumn{2}{|c|}{4975.853}\\
\hline
\hline
\end{tabular}
\caption{Best-fit values and $68\%$ confidence limits for chameleon-type model(and $95\%$ confidence limits in parenthesis for $\beta_1$).}
\label{Tab3}
\end{center}
\end{table*} 

The global analysis results for $f(R)$ gravity can be read in the second, third, fourth and fifth columns of Tab. \ref{Tab2}, which are based on WMAP9yr, {\it Planck} + WP, and {\it Planck} + WP + lensing and {\it Planck} + WP + lensing + BAO + HST + Union2.1 data sets. 

Firstly, we can see that {\it Planck} CMB temperature power-spectrum with WP can give an upper bound of $B_0<0.91$ (hereafter we quote the significance at $95\%$C.L. for modified gravity parameter, such as $B_0$ and $\beta_1$). Compared with the WMAP9yr result, $B_0<3.37$, it improves the upper bound by a factor $3.7$. Secondly, by adding lensing data the results can be further improved by a factor $5.1$ ($B_0<0.18$). Finally, we arrive at our best bound of $B_0<0.12$ by using all data sets. 
In addition, we notice that, due to the degeneracy between $B_0$ and the dark matter density, the {\it Planck} data prefer a slightly lower value of $\Omega_c h^2$ in $f(R)$ model. Consequently, this implies that $f(R)$ gravity favors a slightly larger value of $H_0$.
This can be helpful to relax the tension between {\it Planck} and the other direct measurements of the Hubble parameter, such as that from the HST \cite{Freedman:2000cf}. 
The degenearcy between $B_0$ and $\Omega_c h^2$ is illustrated in Fig.~\ref{3rdpeak}, where it is evident 
that we can fit a lower value of the third peak by increasing $B_0$ while 
keeping $\Omega_c h^2$ fixed.

Marginalized likelihoods for all the parameters are shown in Fig. \ref{MG_1d}. We also highlight the 2D likelihood in the parameter space of $B_0$ and $H_0$ in Fig. \ref{B0_H0} and the marginalized likelihood for $B_0$ in Fig. \ref{B0}. Let us notice that the $B_0$ likelihood from WMAP9yr data (gray curve) has a prominent second peak around $B_0=2.5$. This is due to the non-linear dependence of the ISW effect on $B_0$ in $f(R)$ gravity. 
Since with WMAP resolution the lensing signal is quite weak, the 
main contribution to the $B_0$ constraint in WMAP data comes from the ISW effect. 
As shown in Fig. \ref{isw}, under our parameter value choice (we fix all the other cosmological parameters as the mean values of the {\it Planck} base $\Lambda$CDM model), from $B_0=0$ to $B_0\sim 1$ the slope of the spectrum in the ISW-dominated regime becomes gradually flat and approaches the Sachs-Wolfe plateau. After that, if one continues to increase till $B_0\sim 2$, the power-spectrum will bounce again and get closer to that of general relativity. If one further increases the $B_0$ value, the spectrum curve in the ISW regime will rise up above that of general relativity. Moreover, once we marginalize over all the other cosmological parameters, the turning point $B_0\sim 1$ will shift to around $B_0\sim 1.5$, and the second peak $B_0\sim 2$ moves to $B_0\sim 2.5$.

\begin{figure}[ht]
\begin{center}
  \includegraphics[width=0.5\textwidth]{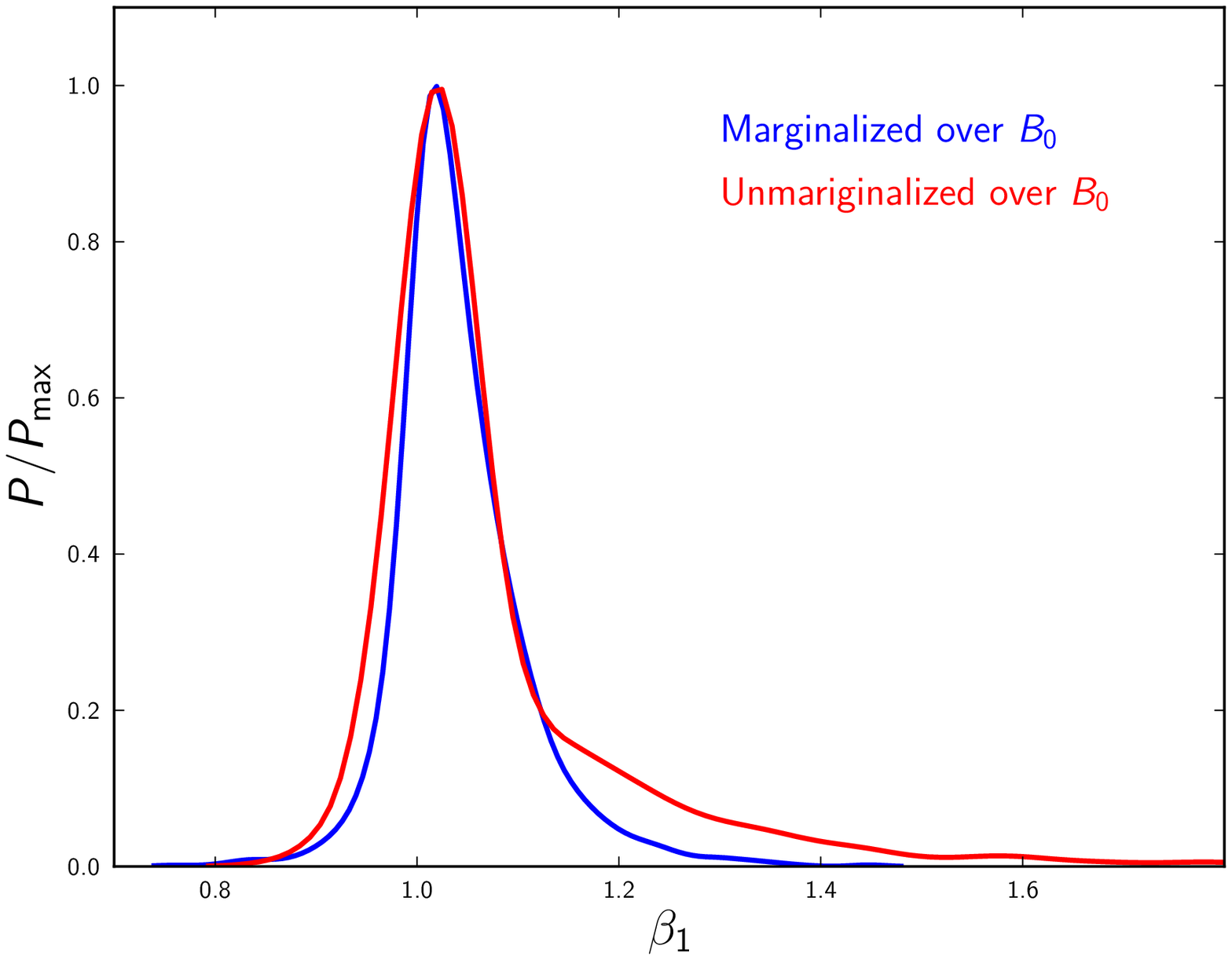}
  \caption{\label{fig:B1} Likelihood of $\beta_1$ with(out) marginalization over $B_0$ by using ${\it Planck}$+WP+Lensing+BAO+HST+Union2 data sets. }
\end{center}
\end{figure}

Compared with $f(R)$ gravity, the chameleon-type model includes the other two free parameters $\beta_1$ and $s$, which are fixed to $4/3$ and $4$ in the former case. 
Due to the amount of extra modified gravity parameters and the degeneracy among them, we find that the {\it Planck} constraints on the parameters $B_0$ and $s$ are still quite 
loose, with no obvious improvement when comparing to WMAP9yr results. However, we are able to improve the constraints on $\beta_1$: we find 
$\beta_1=1.043^{+0.163}_{-0.104}$ at $95\%$C.L. compared with  $\beta_1=0.893^{+0.647}_{-0.695}$ at $95\%$C.L. from WMAP9yr. The detailed global analysis results can be found in Tab. 
\ref{Tab3} and Fig. \ref{CM_1d}. Confidence regions in the $\beta_1$-$H_0$ plane, after marginalizing over the other parameters, are 
shown in Fig. \ref{beta1_H0}. One could notice that the value $\beta_1=4/3$, corresponding to $f(R)$ models, is well outside the $3\sigma$ confidence region. However this does 
not rule out $f(R)$ model by any means given the very loose constraints on the other two relevant $f(R)$ parameters $B_0$ and $s$. 
In Fig. \ref{fig:B1}, we compare the likelihood of $\beta_1$ with(out) marginalization over $B_0$ by using ${\it Planck}$+WP+Lensing+BAO+HST+Union2 data sets.
It clearly shows that the stringent constraint on $\beta_1$ is due to the marginalization effect on $B_0$, whose constraint is very loose for
the chameleon-type model via current data sets. And we have alse tested that if we fix $B_0=0.001$ and use the same data sets, 
the marginalized $2\sigma$ confidence level for $\beta_1$ is $0.971^{+0.700}_{-0.746}$, which reconciles with $f(R)$ gravity very well. 
We can also see in Tab. \ref{Tab3} that the chameleon-type model favors a slightly higher Hubble parameter, for the same reason as explained for $f(R)$ gravity.

\begin{acknowledgments}
NB and BH are indebted to Philippe Brax for useful discussions. 
 We also thank Jason Dossett, Alireza Hojjati and Alessandra Silvestri
for the useful correspondence and discussion of the numerical codes.
BH thanks Zhenhui Zhang for discussion and ITP-CAS for the 
hospitality where some parts of this work are finished. 
The numerical calculations are performed on clusters at ITP-CAS
and INFN-PD. 
The research of N.B., M.L. and S.M. has been partially supported by 
the ASI/INAF Agreement No. I/072/09/0 for the Planck LFI Activity of 
Phase E2.
\end{acknowledgments}



\vspace*{0.2cm}
\begin{thebibliography}{99}

\bibitem{Bertschinger:2006aw} 
  E.~Bertschinger,
  Astrophys.\ J.\  {\bf 648}, 797 (2006)
  [astro-ph/0604485].
  
\bibitem{Song:2006ej}
  Y.~-S.~Song, W.~Hu and I.~Sawicki,
  Phys.\ Rev.\ D {\bf 75}, 044004 (2007)
  [astro-ph/0610532].
  
\bibitem{Brax:2008hh}
  P.~Brax, C.~van de Bruck, A.~-C.~Davis and D.~J.~Shaw,
  Phys.\ Rev.\ D {\bf 78}, 104021 (2008)
  [arXiv:0806.3415 [astro-ph]].
  
\bibitem{Zhao:2009fn} 
  G.~-B.~Zhao, L.~Pogosian, A.~Silvestri and J.~Zylberberg,
  Phys.\ Rev.\ Lett.\  {\bf 103}, 241301 (2009)
  [arXiv:0905.1326 [astro-ph.CO]].

\bibitem{Zhao:2010dz} 
  G.~-B.~Zhao, T.~Giannantonio, L.~Pogosian, A.~Silvestri, D.~J.~Bacon, K.~Koyama, R.~C.~Nichol and Y.~-S.~Song,
  Phys.\ Rev.\ D {\bf 81}, 103510 (2010)
  [arXiv:1003.0001 [astro-ph.CO]].
  
\bibitem{Hojjati:2011xd} 
  A.~Hojjati, G.~-B.~Zhao, L.~Pogosian, A.~Silvestri, R.~Crittenden and K.~Koyama,
  Phys.\ Rev.\ D {\bf 85}, 043508 (2012)
  [arXiv:1111.3960 [astro-ph.CO]].

\bibitem{Bertschinger:2008zb}
  E.~Bertschinger and P.~Zukin,
  Phys.\ Rev.\ D {\bf 78}, 024015 (2008)
  [arXiv:0801.2431 [astro-ph]].

\bibitem{Brax:2012gr}
  P.~Brax, A.~-C.~Davis, B.~Li and H.~A.~Winther,
  Phys.\ Rev.\ D {\bf 86}, 044015 (2012)
  [arXiv:1203.4812 [astro-ph.CO]].
  
\bibitem{Zuntz:2011aq}
  J.~Zuntz, T.~Baker, P.~Ferreira and C.~Skordis,
  arXiv:1110.3830 [astro-ph.CO].

\bibitem{Hojjati:2012rf}
  A.~Hojjati, L.~Pogosian, A.~Silvestri and S.~Talbot,
  arXiv:1210.6880 [astro-ph.CO].

\bibitem{Hu:2007pj}
  W.~Hu and I.~Sawicki,
  Phys.\ Rev.\ D {\bf 76}, 104043 (2007)
  [arXiv:0708.1190 [astro-ph]].
 
\bibitem{Fang:2008sn}
  W.~Fang, W.~Hu and A.~Lewis,
  Phys.\ Rev.\ D {\bf 78}, 087303 (2008)
  [arXiv:0808.3125 [astro-ph]].
  
\bibitem{Lombriser:2013aj}
  L.~Lombriser, J.~Yoo and K.~Koyama,
  arXiv:1301.3132 [astro-ph.CO].

\bibitem{Baker:2011jy}
  T.~Baker, P.~G.~Ferreira, C.~Skordis and J.~Zuntz,
  Phys.\ Rev.\ D {\bf 84}, 124018 (2011)
  [arXiv:1107.0491 [astro-ph.CO]].

\bibitem{Baker:2012zs}
  T.~Baker, P.~G.~Ferreira and C.~Skordis,
  Phys.\ Rev.\ D {\bf 87}, 024015 (2013)
  [arXiv:1209.2117 [astro-ph.CO]].

\bibitem{ArkaniHamed:2003uy} 
  N.~Arkani-Hamed, H.~-C.~Cheng, M.~A.~Luty and S.~Mukohyama,
  JHEP {\bf 0405}, 074 (2004)
  [hep-th/0312099].

\bibitem{Gubitosi:2012hu}
  G.~Gubitosi, F.~Piazza and F.~Vernizzi,
  arXiv:1210.0201 [hep-th].

\bibitem{Bloomfield:2012ff}
  J.~K.~Bloomfield, E.~E.~Flanagan, M.~Park and S.~Watson,
  arXiv:1211.7054 [astro-ph.CO].
  
\bibitem{Creminelli:2006xe}
  P.~Creminelli, M.~A.~Luty, A.~Nicolis and L.~Senatore,
  JHEP {\bf 0612}, 080 (2006)
  [hep-th/0606090].

\bibitem{Cheung:2007st}
  C.~Cheung, P.~Creminelli, A.~L.~Fitzpatrick, J.~Kaplan and L.~Senatore,
  JHEP {\bf 0803}, 014 (2008)
  [arXiv:0709.0293 [hep-th]].
  
\bibitem{Creminelli:2008wc} 
  P.~Creminelli, G.~D'Amico, J.~Norena and F.~Vernizzi,
  JCAP {\bf 0902}, 018 (2009)
  [arXiv:0811.0827 [astro-ph]].
  
\bibitem{Gleyzes:2013ooa} 
  J.~Gleyzes, D.~Langlois, F.~Piazza and F.~Vernizzi,
  arXiv:1304.4840 [hep-th].
  
\bibitem{Bloomfield:2013efa} 
  J.~Bloomfield,
  arXiv:1304.6712 [astro-ph.CO].

\bibitem{Sachs:1967er} 
  R.~K.~Sachs and A.~M.~Wolfe,
  Astrophys.\ J.\  {\bf 147}, 73 (1967)
  [Gen.\ Rel.\ Grav.\  {\bf 39}, 1929 (2007)].
  
\bibitem{Zhang:2005vt}
  P.~Zhang,
  ``Testing $f(R)$ gravity against the large scale structure of the universe.,''
  Phys.\ Rev.\ D {\bf 73}, 123504 (2006)
  [astro-ph/0511218].
  
\bibitem{Ho:2008bz} 
  S.~Ho, C.~Hirata, N.~Padmanabhan, U.~Seljak and N.~Bahcall,
  Phys.\ Rev.\ D {\bf 78}, 043519 (2008)
  [arXiv:0801.0642 [astro-ph]].
  
\bibitem{Daniel:2010ky} 
  S.~F.~Daniel, E.~V.~Linder, T.~L.~Smith, R.~R.~Caldwell, A.~Cooray, A.~Leauthaud and L.~Lombriser,
  Phys.\ Rev.\ D {\bf 81}, 123508 (2010)
  [arXiv:1002.1962 [astro-ph.CO]].
    
\bibitem{Marchini:2013lpp} 
  A.~Marchini, A.~Melchiorri, V.~Salvatelli and L.~Pagano,
  Phys.\ Rev.\ D {\bf 87}, 083527 (2013)
  [arXiv:1302.2593 [astro-ph.CO]].
      
\bibitem{Bean:2010zq} 
  R.~Bean and M.~Tangmatitham,
  Phys.\ Rev.\ D {\bf 81}, 083534 (2010)
  [arXiv:1002.4197 [astro-ph.CO]].
  
\bibitem{Yamamoto:2006yv} 
  K.~Yamamoto, B.~A.~Bassett, R.~C.~Nichol and Y.~Suto,
  Phys.\ Rev.\ D {\bf 74}, 063525 (2006)
  [astro-ph/0605278].
  
\bibitem{Said:2013jxa} 
  N.~Said, C.~Baccigalupi, M.~Martinelli, A.~Melchiorri and A.~Silvestri,
  arXiv:1303.4353 [astro-ph.CO].
  
\bibitem{DiValentino:2012yg} 
  E.~Di Valentino, A.~Melchiorri, V.~Salvatelli and A.~Silvestri,
  Phys.\ Rev.\ D {\bf 86}, 063517 (2012)
  [arXiv:1204.5352 [astro-ph.CO]].
  
\bibitem{Hu:2012td} 
  B.~Hu, M.~Liguori, N.~Bartolo and S.~Matarrese,
  arXiv:1211.5032 [astro-ph.CO].
  
\bibitem{Giannantonio:2009gi}
  T.~Giannantonio, M.~Martinelli, A.~Silvestri and A.~Melchiorri,
  JCAP {\bf 1004}, 030 (2010)
  [arXiv:0909.2045 [astro-ph.CO]].
  
\bibitem{Dossett:2011zp} 
  J.~Dossett, J.~Moldenhauer and M.~Ishak,
  Phys.\ Rev.\ D {\bf 84}, 023012 (2011)
  [arXiv:1103.1195 [astro-ph.CO]].
    
\bibitem{Song:2007da} 
  Y.~-S.~Song, H.~Peiris and W.~Hu,
  Phys.\ Rev.\ D {\bf 76}, 063517 (2007)
  [arXiv:0706.2399 [astro-ph]].
  
\bibitem{Jain:2007yk} 
  B.~Jain and P.~Zhang,
  Phys.\ Rev.\ D {\bf 78}, 063503 (2008)
  [arXiv:0709.2375 [astro-ph]].
  
\bibitem{Lombriser:2010mp}
  L.~Lombriser, A.~Slosar, U.~Seljak and W.~Hu,
  arXiv:1003.3009 [astro-ph.CO].
  
\bibitem{Ferraro:2010gh} 
  S.~Ferraro, F.~Schmidt and W.~Hu,
  Phys.\ Rev.\ D {\bf 83}, 063503 (2011)
  [arXiv:1011.0992 [astro-ph.CO]].
  
\bibitem{Lombriser:2013wta} 
  L.~Lombriser, B.~Li, K.~Koyama and G.~-B.~Zhao,
  Phys.\  Rev.\  D 87, {\bf 123511} (2013)
  [arXiv:1304.6395 [astro-ph.CO]].

\bibitem{Li:2012by} 
  B.~Li, W.~A.~Hellwing, K.~Koyama, G.~-B.~Zhao, E.~Jennings and C.~M.~Baugh,
  arXiv:1206.4317 [astro-ph.CO].
  
\bibitem{Asaba:2013xql} 
  S.~Asaba, C.~Hikage, K.~Koyama, G.~-B.~Zhao, A.~Hojjati and L.~Pogosian,
  arXiv:1306.2546 [astro-ph.CO].
  
\bibitem{Thomas:2008tp} 
  S.~A.~Thomas, F.~B.~Abdalla and J.~Weller,
  Mon.\ Not.\ Roy.\ Astron.\ Soc.\  {\bf 395}, 197 (2009)
  [arXiv:0810.4863 [astro-ph]].
  
\bibitem{Cai:2011wj} 
  Y.~-C.~Cai and G.~Bernstein,
  Mon.\ Not.\ Roy.\ Astron.\ Soc.\  {\bf 422}, 1045 (2012)
  [arXiv:1112.4478 [astro-ph.CO]].
  
\bibitem{Jennings:2012pt} 
  E.~Jennings, C.~M.~Baugh, B.~Li, G.~-B.~Zhao and K.~Koyama,
  arXiv:1205.2698 [astro-ph.CO].
  
\bibitem{Raccanelli:2012gt} 
  A.~Raccanelli, D.~Bertacca, D.~Pietrobon, F.~Schmidt, L.~Samushia, N.~Bartolo, O.~Dore and S.~Matarrese {\it et al.},
  arXiv:1207.0500 [astro-ph.CO].
  
\bibitem{Zhang:2007nk} 
  P.~Zhang, M.~Liguori, R.~Bean and S.~Dodelson,
  Phys.\ Rev.\ Lett.\  {\bf 99}, 141302 (2007)
  [arXiv:0704.1932 [astro-ph]].
  
\bibitem{Hirata:2008cb} 
  C.~M.~Hirata, S.~Ho, N.~Padmanabhan, U.~Seljak and N.~A.~Bahcall,
  Phys.\ Rev.\ D {\bf 78}, 043520 (2008)
  [arXiv:0801.0644 [astro-ph]].
  
\bibitem{Zhao:2008bn}
  G.~-B.~Zhao, L.~Pogosian, A.~Silvestri and J.~Zylberberg,
  Phys.\ Rev.\ D {\bf 79}, 083513 (2009)
  [arXiv:0809.3791 [astro-ph]].
  
\bibitem{Laszlo:2011sv} 
  I.~Laszlo, R.~Bean, D.~Kirk and S.~Bridle,
  arXiv:1109.4535 [astro-ph.CO].
  
\bibitem{Tereno:2010dt} 
  I.~Tereno, E.~Semboloni and T.~Schrabback,
  Astron.\ Astrophys.\  {\bf 530}, A68 (2011)
  [arXiv:1012.5854 [astro-ph.CO]].
  
\bibitem{Reyes:2010tr} 
  R.~Reyes, R.~Mandelbaum, U.~Seljak, T.~Baldauf, J.~E.~Gunn, L.~Lombriser and R.~E.~Smith,
  Nature {\bf 464}, 256 (2010)
  [arXiv:1003.2185 [astro-ph.CO]].
  
\bibitem{Wang:2010ug} 
  X.~Wang, X.~Chen and C.~Park,
  Astrophys.\ J.\  {\bf 747}, 48 (2012)
  [arXiv:1010.3035 [astro-ph.CO]].
    
\bibitem{Hall:2012wd} 
  A.~Hall, C.~Bonvin and A.~Challinor,
  arXiv:1212.0728 [astro-ph.CO].
  
\bibitem{Yamamoto:2010ie} 
  K.~Yamamoto, G.~Nakamura, G.~Hutsi, T.~Narikawa and T.~Sato,
  Phys.\ Rev.\ D {\bf 81}, 103517 (2010)
  [arXiv:1004.3231 [astro-ph.CO]].
    
\bibitem{He:2012wq} 
  J.~-h.~He,
  Phys.\ Rev.\ D {\bf 86}, 103505 (2012)
  [arXiv:1207.4898 [astro-ph.CO]].
  
\bibitem{GilMarin:2011xq} 
  H.~Gil-Marin, F.~Schmidt, W.~Hu, R.~Jimenez and L.~Verde,
  JCAP {\bf 1111}, 019 (2011)
  [arXiv:1109.2115 [astro-ph.CO]].
  
\bibitem{Bartolo:2013ws} 
  N.~Bartolo, E.~Bellini, D.~Bertacca and S.~Matarrese,
  JCAP {\bf 1303}, 034 (2013)
  [arXiv:1301.4831 [astro-ph.CO]].
  
\bibitem{Zhao:2010qy} 
  G.~-B.~Zhao, B.~Li and K.~Koyama,
  Phys.\ Rev.\ D {\bf 83}, 044007 (2011)
  [arXiv:1011.1257 [astro-ph.CO]].
  
\bibitem{He:2013vwa} 
  J.~-h.~He, B.~Li and Y.~Jing,
  arXiv:1305.7333 [astro-ph.CO].
  
\bibitem{Li:2010zw} 
  B.~Li, D.~F.~Mota and J.~D.~Barrow,
  Astrophys.\ J.\  {\bf 728}, 109 (2011)
  [arXiv:1009.1400 [astro-ph.CO]].
  
\bibitem{Ade:2013ktc} 
  P.~A.~R.~Ade {\it et al.}  [Planck Collaboration],
  arXiv:1303.5062 [astro-ph.CO].
  
\bibitem{Planck:2013kta} 
  P.~A.~R.~Ade {\it et al.}  [Planck Collaboration],
  arXiv:1303.5075 [astro-ph.CO].

\bibitem{Ade:2013tyw} 
  P.~A.~R.~Ade {\it et al.}  [Planck Collaboration],
  arXiv:1303.5077 [astro-ph.CO].
  
\bibitem{Ade:2013ydc} 
  P.~A.~R.~Ade {\it et al.}  [Planck Collaboration],
  arXiv:1303.5084 [astro-ph.CO].
  
\bibitem{Ade:2013dsi} 
  P.~A.~R.~Ade {\it et al.}  [Planck Collaboration],
  arXiv:1303.5079 [astro-ph.CO].
  
\bibitem{Li:2013nwa} 
  Y.~-C.~Li, F.~-Q.~Wu and X.~Chen,
  arXiv:1305.0055 [astro-ph.CO].
  
\bibitem{Li:2013dha} 
  M.~Li, X.~-D.~Li, Y.~-Z.~Ma, X.~Zhang and Z.~Zhang,
  arXiv:1305.5302 [astro-ph.CO].
  
\bibitem{Pettorino:2013oxa} 
  V.~Pettorino,
  arXiv:1305.7457 [astro-ph.CO].

\bibitem{Salvatelli:2013wra} 
  V.~Salvatelli and A.~Marchini,
  arXiv:1304.7119 [astro-ph.CO].

\bibitem{Marchini:2013oya} 
  A.~Marchini and V.~Salvatelli,
  arXiv:1307.2002 [astro-ph.CO].

\bibitem{He:2013qha} 
  J.~-h.~He,
  arXiv:1307.4876 [astro-ph.CO].
      
\bibitem{Mota:2006fz} 
  D.~F.~Mota and D.~J.~Shaw,
  Phys.\ Rev.\ D {\bf 75}, 063501 (2007)
  [hep-ph/0608078].
  
\bibitem{Brax:2004qh} 
  P.~Brax, C.~van de Bruck, A.~-C.~Davis, J.~Khoury and A.~Weltman,
  Phys.\ Rev.\ D {\bf 70}, 123518 (2004)
  [astro-ph/0408415].
    
\bibitem{Pietroni:2005pv} 
  M.~Pietroni,
  Phys.\ Rev.\ D {\bf 72}, 043535 (2005)
  [astro-ph/0505615].
  
\bibitem{Olive:2007aj} 
  K.~A.~Olive and M.~Pospelov,
  Phys.\ Rev.\ D {\bf 77}, 043524 (2008)
  [arXiv:0709.3825 [hep-ph]].
  
\bibitem{Hinterbichler:2010es} 
  K.~Hinterbichler and J.~Khoury,
  Phys.\ Rev.\ Lett.\  {\bf 104}, 231301 (2010)
  [arXiv:1001.4525 [hep-th]].
  
\bibitem{Brax:2010gi} 
  P.~Brax, C.~van de Bruck, A.~-C.~Davis and D.~Shaw,
  Phys.\ Rev.\ D {\bf 82}, 063519 (2010)
  [arXiv:1005.3735 [astro-ph.CO]].
  
\bibitem{Hojjati:2011ix}
  A.~Hojjati, L.~Pogosian and G.~-B.~Zhao,
  JCAP {\bf 1108}, 005 (2011)
  [arXiv:1106.4543 [astro-ph.CO]].
  
\bibitem{Lewis:1999bs}
  A.~Lewis, A.~Challinor and A.~Lasenby,
  Astrophys.\ J.\  {\bf 538}, 473 (2000)
  [astro-ph/9911177].
  
\bibitem{Bennett:2012fp} 
  C.~L.~Bennett, D.~Larson, J.~L.~Weiland, N.~Jarosik, G.~Hinshaw, N.~Odegard, K.~M.~Smith and R.~S.~Hill {\it et al.},
  arXiv:1212.5225 [astro-ph.CO].
  
\bibitem{Hinshaw:2012aka} 
  G.~Hinshaw {\it et al.}  [WMAP Collaboration],
  arXiv:1212.5226 [astro-ph.CO].
  
\bibitem{DeFelice:2010aj}
  A.~De Felice and S.~Tsujikawa,
  Living Rev.\ Rel.\  {\bf 13}, 3 (2010)
  [arXiv:1002.4928 [gr-qc]].

\bibitem{Lewis:2002ah} 
  A.~Lewis and S.~Bridle,
  Phys.\ Rev.\ D {\bf 66}, 103511 (2002)
  [astro-ph/0205436].

\bibitem{Ade:2013zuv} 
  P.~A.~R.~Ade {\it et al.}  [Planck Collaboration],
  arXiv:1303.5076 [astro-ph.CO].
  
\bibitem{Beutler:2011hx} 
  F.~Beutler, C.~Blake, M.~Colless, D.~H.~Jones, L.~Staveley-Smith, L.~Campbell, Q.~Parker and W.~Saunders {\it et al.},
  Mon.\ Not.\ Roy.\ Astron.\ Soc.\  {\bf 416}, 3017 (2011)
  [arXiv:1106.3366 [astro-ph.CO]].
  
\bibitem{Percival:2009xn} 
  W.~J.~Percival {\it et al.}  [SDSS Collaboration],
  Mon.\ Not.\ Roy.\ Astron.\ Soc.\  {\bf 401}, 2148 (2010)
  [arXiv:0907.1660 [astro-ph.CO]].
  
\bibitem{Padmanabhan:2012hf} 
  N.~Padmanabhan, X.~Xu, D.~J.~Eisenstein, R.~Scalzo, A.~J.~Cuesta, K.~T.~Mehta and E.~Kazin,
  Mon.\ Not.\ Roy.\ Astron.\ Soc.\  {\bf 427}, no. 3, 2132 (2012)
  [arXiv:1202.0090 [astro-ph.CO]].
  
\bibitem{Anderson:2012sa} 
  L.~Anderson, E.~Aubourg, S.~Bailey, D.~Bizyaev, M.~Blanton, A.~S.~Bolton, J.~Brinkmann and J.~R.~Brownstein {\it et al.},
  Mon.\ Not.\ Roy.\ Astron.\ Soc.\  {\bf 427}, no. 4, 3435 (2013)
  [arXiv:1203.6594 [astro-ph.CO]].
  
\bibitem{Freedman:2000cf} 
  W.~L.~Freedman {\it et al.}  [HST Collaboration],
  Astrophys.\ J.\  {\bf 553}, 47 (2001)
  [astro-ph/0012376].
    
    
\bibitem{Riess:2011yx} 
  A.~G.~Riess, L.~Macri, S.~Casertano, H.~Lampeitl, H.~C.~Ferguson, A.~V.~Filippenko, S.~W.~Jha and W.~Li {\it et al.},
  Astrophys.\ J.\  {\bf 730}, 119 (2011)
  [Erratum-ibid.\  {\bf 732}, 129 (2011)]
  [arXiv:1103.2976 [astro-ph.CO]].
  
\bibitem{Suzuki:2011hu} 
  N.~Suzuki, D.~Rubin, C.~Lidman, G.~Aldering, R.~Amanullah, K.~Barbary, L.~F.~Barrientos and J.~Botyanszki {\it et al.},
  Astrophys.\ J.\  {\bf 746}, 85 (2012)
  [arXiv:1105.3470 [astro-ph.CO]].
  

  
 


  



\end{thebibliography}
\end{document}